\documentclass[aps,prl,twocolumn,superscriptaddress,nofootinbib]{revtex4}

\usepackage{graphicx}
\usepackage{dcolumn}
\usepackage{bm}
\usepackage{amssymb}
\usepackage{amsfonts}
\usepackage{latexsym}
\usepackage{enumerate}
\usepackage[dvipdfmx]{color} 
\usepackage{amsmath}
\usepackage[dvipdfmx]{epsfig}
\usepackage{times}
\usepackage{algorithm}
\usepackage{algorithmic}
\usepackage{cases}

\begin{document}
\widetext
\title{Catalytic Transformation from Computationally Universal to Strictly Universal Measurement-Based Quantum Computation}
\author{Yuki Takeuchi}
\email{yuki.takeuchi@ntt.com}
\affiliation{NTT Communication Science Laboratories, NTT Corporation, 3-1 Morinosato Wakamiya, Atsugi, Kanagawa 243-0198, Japan}
\affiliation{NTT Research Center for Theoretical Quantum Information, NTT Corporation, 3-1 Morinosato Wakamiya, Atsugi, Kanagawa 243-0198, Japan}

\begin{abstract}
There are two types of universality in measurement-based quantum computation (MBQC): {\it strict} and {\it computational}.
It is well known that the former is stronger than the latter.
We present a method of transforming from a certain type of computationally universal MBQC to a strictly universal one.
Our method simply replaces a single qubit in a resource state with a Pauli-$Y$ eigenstate.
We applied our method to show that hypergraph states can be made strictly universal with only Pauli measurements, while only computationally universal hypergraph states were known.
\end{abstract}
\maketitle

Quantum computers solve several problems faster than classical computers with the best known classical algorithms~\cite{S97,WZ06,S06}.
Driven by this advantage, tremendous effort has been devoted to developing quantum computers, and several quantum-computing models were proposed such as quantum circuit model~\cite{NC10}, measurement-based quantum computation (MBQC)~\cite{RB01,RBB03}, and adiabatic quantum computation~\cite{ADKLLR07}.
Although these models have unique features, they are the same in terms of computational capability (i.e., what problems can be solved in polynomial time).
More concretely, these models can execute ``any" quantum computation, hence are called universal quantum-computing models.

There are two types of universality in quantum computation~\cite{A03}.
One is {\it strict} universality, which is the strongest notion of universality.
It means that any unitary operator can be implemented with an arbitrary high accuracy; hence, any quantum state can also be generated.
However, a restricted class of unitary operators is sufficient to generate the output probability distribution of any quantum circuit with an arbitrary high accuracy~\cite{S02,A03}.
Therefore, we can define a weaker notion of universality called {\it computational} universality.
To clarify the difference between these two notions, let us consider the quantum circuit model with $n$ initialized input qubits $|0^n\rangle$.
In this model, the universality is determined by gate sets; $\{H, T, \Lambda(Z)\}$ and $\{H, \Lambda(S)\}$ are examples of strictly universal gate sets.
Here, $H\equiv|+\rangle\langle 0|+|-\rangle\langle 1|$, where $|\pm\rangle\equiv(|0\rangle\pm|1\rangle)/\sqrt{2}$, is the Hadamard gate, $T\equiv|0\rangle\langle 0|+e^{i\pi/4}|1\rangle\langle 1|$ is the $T$ gate, $\Lambda(U)\equiv|0\rangle\langle 0|\otimes I+|1\rangle\langle 1|\otimes U$ is the controlled-$U$ gate for any single-qubit unitary operator $U$, $I$ is the two-dimensional identity gate, $Z\equiv T^4$ is the Pauli-$Z$ gate, and $S\equiv T^2$ is the $S$ gate.
Real unitary operators, however, are sufficient to construct the computationally universal gate set $\{H, CCZ\}$~\cite{FN1}, where
$CCZ\equiv I^{\otimes 3}-2|111\rangle\langle 111|$ is the controlled-controlled-$Z$ ($CCZ$) gate.
By definition, it is trivial that strictly universal gate sets are also computationally universal, but the opposite does not hold.
The computationally universal gate set $\{H, CCZ\}$ is insufficient to generate complex quantum states, the amplitudes of which include imaginary numbers.
For example, the $n$-qubit quantum state $|\psi_t\rangle=(|0^n\rangle+i|1^n\rangle)/\sqrt{2}$ cannot be generated with fidelity larger than $1/2$.
This is because any quantum state generated by applying $H$ and $CCZ$ gates to $|0^n\rangle$ is written as a real quantum state $|\phi_r\rangle=\sum_{z\in\{0,1\}^n}c_z|z\rangle$ with real numbers $\{c_z\}_{z\in\{0,1\}^n}$ satisfying $\sum_{z\in\{0,1\}^n}c_z^2=1$; hence, $|\langle\psi_t|\phi_r\rangle|^2=(c_{0^n}^2+c_{1^n}^2)/2\le1/2$.

The difference between complex and real quantum states becomes more apparent when we focus on multiparty quantum information processing.
There exists a task conducted by three parties that can be achieved by using complex quantum states but cannot by using real ones~\cite{RTWLTGAN21}, and their difference was already experimentally observed by using photons~\cite{LMWTCFYRTLGANWF22} and superconducting qubits~\cite{CWLWYSWGDLZPZCLP22}.
Given the importance of complex quantum states, the resource theory of imaginarity has also been developed~\cite{HG18,WKRSXLGS21L,WKRSXLGS21A,WKSRXLGS23}.
These results would indicate the necessity of strict universality.

As mentioned above, MBQC is a universal quantum-computing model~\cite{RB01}.
It proceeds by adaptively measuring qubits of a resource state one by one.
Its universality is determined by a given resource state (and available measurement bases).
For both types of universality, several resource states were proposed.
Cluster states~\cite{BR01}, Affleck-Kennedy-Lieb-Tasaki (AKLT) states~\cite{WAR11}, and parity-phase graph states~\cite{KW19}, which are weighted graph states~\cite{DHHLB05,FN3}, are common examples of strictly universal resource states.
For computational universality, several hypergraph states were found, which require only Pauli measurements~\cite{MM16,MM18,GGM19,TMH19,YFTTK20}.
As with the quantum circuit model, strictly universal resource states are trivially computationally universal, but the opposite is not true.
Despite the importance of understanding the hardness of developing strictly universal quantum computers, the gap between these two types of universality has been less explored in MBQC.

In this Letter, we present a method of transforming to the strictly universal MBQC from any computationally universal one that can precisely implement $H$ and $CCZ$.
Our method is quite simple in that it simply replaces a qubit in a resource state of a computationally universal MBQC with a Pauli-$Y$ eigenstate $|+i\rangle\equiv(|0\rangle+i|1\rangle)/\sqrt{2}$.
Since the added $|+i\rangle$ works like a catalyst in chemistry, we call the transformation with our method catalytic transformation.
In fact, $|+i\rangle$ achieves a strictly universal MBQC, while it is invariant during MBQC [see Fig.~\ref{Sapply} (b)].
An advantage of our method is that the required measurement bases are the same before and after the transformation.
To devise our method, we first show that $|1\rangle$ can be deterministically prepared by applying $H$ and $CCZ$ gates to $|000\rangle$.
We then show that $S$ is deterministically applicable to any quantum state $|\psi\rangle$ by applying $H$ and $CCZ$ gates to $|1\rangle|+i\rangle|\psi\rangle$.
As an important point, $|+i\rangle$ is not consumed when we implement the $S$ gate; hence, it can be repeatedly used to apply multiple $S$ gates.
This approach can be considered as an example of the catalytic embeddings introduced in Ref.~\cite{ACGMMR23}, i.e., a catalytic embedding of $S$ over $\{H,CCZ\}$.
Thus, our results would exhibit the usefulness of the catalytic embeddings for MBQC.
Another example was given in Ref.~\cite{AGKMMR23}, which generalizes the representation of complex numbers used in Ref.~\cite{A03}.
Our approach is also related to conversions with catalyst states~\cite{BCHK20}.
In summary, by using our method, we achieve strictly universal MBQC that can execute any quantum computation composed of $\{H,S,CCZ\}$.

A weakness of our method is that the $S$ gates cannot be applied in parallel because each $S$ requires a single $|+i\rangle$, but the transformed resource state includes only a single $|+i\rangle$.
Toward relaxing this weakness, we also propose a technique of duplicating $|+i\rangle$ in Appendix C.
Its construction is inspired by the $|T\rangle$-catalyzed $|CCZ\rangle\rightarrow2|T\rangle$ factory (C2T factory)~\cite{GF19}, which is a technique in quantum error correction.
Here, $|T\rangle\equiv T|+\rangle$ and $|CCZ\rangle\equiv CCZ(|+\rangle^{\otimes 3})$ are magic states~\cite{BK05} for the $T$ and $CCZ$ gates, respectively.
The C2T factory outputs $|T\rangle^{\otimes 3}$ by applying Clifford gates to $|CCZ\rangle|T\rangle$.
In keeping with the terminologies in previous studies~\cite{JP99,C11}, the third output qubit is called a catalyst.
As an interesting point, the third output qubit can be used as an input of the next C2T factory; hence, we can reinterpret it as a method of generating $|T\rangle^{\otimes 2}$ from $|CCZ\rangle$.
We propose a similar technique for the magic state $|+i\rangle$ of the $S$ gate.
We generate $|+i\rangle^{\otimes 2}$ by applying $H$ and $CCZ$ gates to $|1\rangle|0\rangle|+i\rangle$ and can similarly use the second output qubit as an input of the next duplication.

To concretely reveal the usefulness of our transformation method, we apply it to the MBQC with the hypergraph state in Ref.~\cite{TMH19}.
The MBQC in Ref.~\cite{TMH19} achieves computationally universal quantum computation with only Pauli-$X$ and -$Z$ basis measurements.
In any MBQC with hypergraph states, $|+i\rangle$ can be prepared by a measurement in the Pauli-$Y$ basis.
Therefore, our transformation shows that there are strictly universal hypergraph states with measurements in the Pauli-$X$, -$Y$, and -$Z$ bases. To the best of our knowledge, only computationally universal hypergraph states have been known for Pauli measurements.

\begin{figure}[t]
\includegraphics[width=9cm, clip]{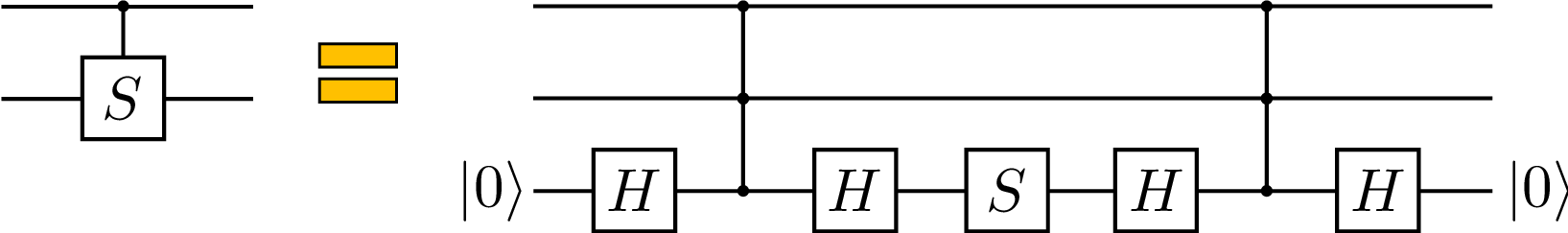}
\caption{Decomposition of $\Lambda(S)$ in terms of $H$, $S$, and $CCZ$ gates. Single ancillary qubit $|0\rangle$ input into the right quantum circuit returns to $|0\rangle$ at the end of the quantum circuit.}
\label{CS}
\end{figure}

\begin{figure}[t]
\includegraphics[width=8cm, clip]{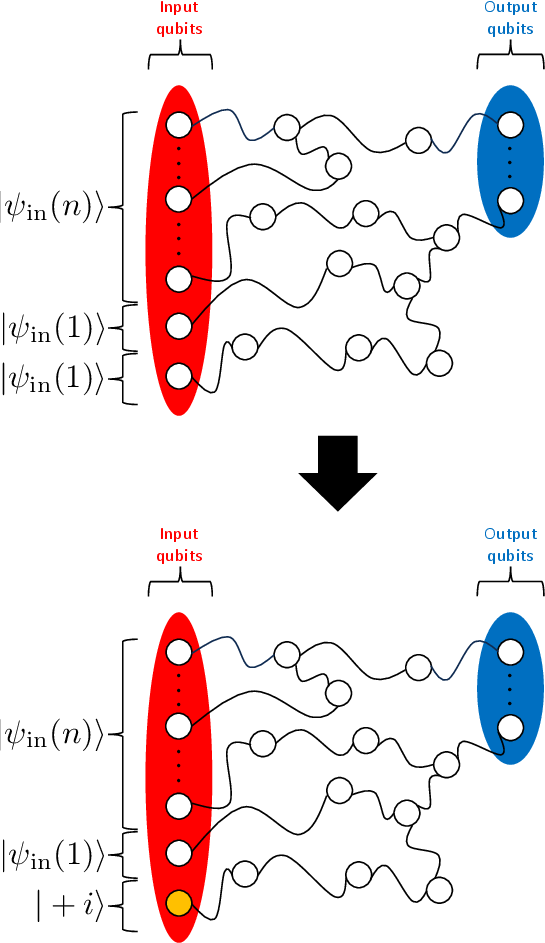}
\caption{Schematic of our transformation. Each circle represents a qubit. Red and blue ellipses represent ${\mathcal C}_I$ and ${\mathcal C}_O$, respectively. Other qubits (not covered by an ellipse) correspond to ${\mathcal C}_M$. By replacing the input qubit (i.e., $|\psi_{\rm in}(1)\rangle$) with $|+i\rangle$ in a given computationally universal resource state $|\Psi_N\rangle$, it becomes strictly universal. Another $|\psi_{\rm in}(1)\rangle$ will be used to prepare $|1\rangle$. Note that this transformation is applicable to any resource state that can precisely implement $H$ and $CCZ$.}
\label{Ours}
\end{figure}

{\it Strictly universal quantum computation with $\{H,S,CCZ\}$.}---As a preliminary to our main results, we show that the gate set $\{H,S,CCZ\}$ is sufficient for strictly universal quantum computation.
A set ${\mathcal G}$ of quantum gates is called strictly universal if there is a positive constant $n_0$ such that the subgroup of unitary operators generated by ${\mathcal G}$ is dense in the special unitary group ${\rm SU}(2^n)$ for any natural number $n\ge n_0$~\cite{A03}.
Simply speaking, by combining quantum gates in a strictly universal gate set, any unitary operator can be constructed with an arbitrary high accuracy.
Kitaev showed $\{H,\Lambda(S)\}$ to be a strictly universal gate set with $n_0=2$~\cite{K97}.
With this fact in mind, it is sufficient for our purpose to give a decomposition of $\Lambda(S)$ in terms of $H$, $S$, and $CCZ$ gates.
We give the decomposition in Fig.~\ref{CS} (for the proof, see Appendix A).

{\it Main results.}---Resource states for a MBQC consist of three sections: input section ${\mathcal C}_I$, body ${\mathcal C}_M$, and output section ${\mathcal C}_O$.
This division was introduced for graph states~\cite{RBB03}, but we do not assume that the resource states are graph states.
In fact, our argument holds even for undiscovered computationally universal states that may not be graph states.
For any natural number $n$, let $V_n$ be any $n$-qubit unitary operator composed of $\{H,CCZ\}$.
In this Letter, we call the MBQC computationally universal if and only if for any $n\ge n_0$, $V_n$ can be applied on ${\mathcal C}_O$ by measuring all qubits in ${\mathcal C}_I\cup{\mathcal C}_M$ one by one in appropriate bases.

Let $|\Psi_n\rangle$ be a computationally universal resource state with $n$ input qubits $|\psi_{\rm in}(n)\rangle\equiv(\bigotimes_{i=1}^nU_{\rm in}^{(i)})|0^n\rangle$, where the single-qubit unitary operator $U_{\rm in}^{(i)}$ is $I$ or $H$ for each $1\le i\le n$.
Our purpose is to transform $|\Psi_n\rangle$ to a strictly universal resource state that deterministically implements any unitary operator composed of $\{H,S,CCZ\}$ (up to a byproduct).
To this end, we first expand the size from $n$ to $N=n+2$ regardless of the number of the $S$ gates to be applied.
We then replace a single qubit in ${\mathcal C}_I$ of $|\Psi_N\rangle$ with $|+i\rangle$, as shown in Fig.~\ref{Ours}~\cite{FN2}.
By using this transformed resource state, strictly universal quantum computation is executed on the first $n$ input qubits $|\psi_{\rm in}(n)\rangle$ with the aid of the two ancillary qubits $|\psi_{\rm in}(1)\rangle|+i\rangle$.

\begin{figure}[t]
\includegraphics[width=9cm, clip]{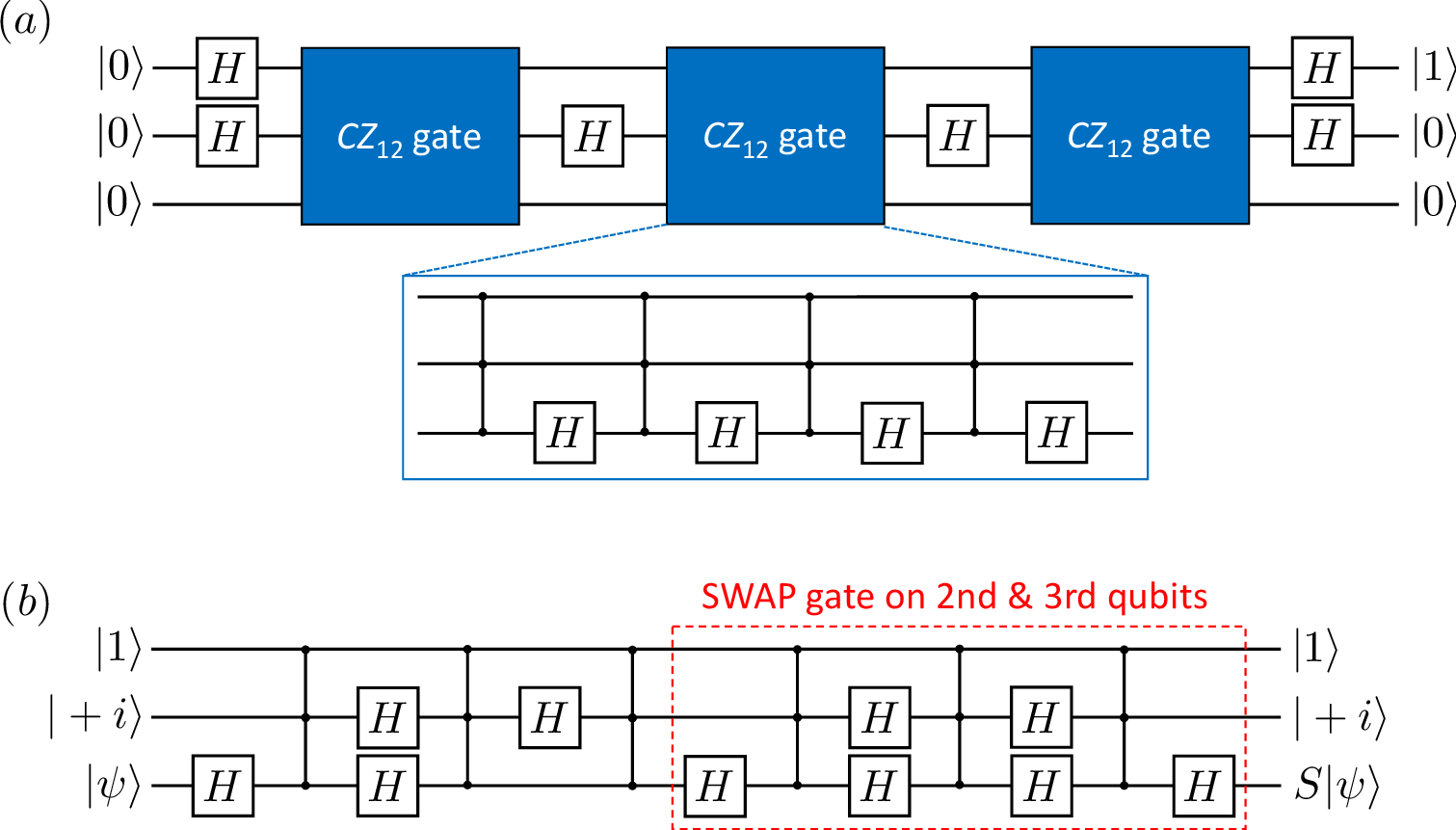}
\caption{Quantum circuits used to implement the $S$ gate. It is important that necessary quantum gates are only $H$ and $CCZ$ gates. (a) Bit flip on $|0\rangle$ can be implemented using only $H$ and $CCZ$ gates. ``$CZ_{12}$ gate'' means that the quantum circuit enclosed by a blue line is equivalent to $\Lambda(Z)\otimes I$. (b) $S$ can be applied to any single-qubit state $|\psi\rangle$ by using $|1\rangle|+i\rangle$ as a catalyst. $|+i\rangle$ is given as an input qubit due to our transformation, and $|1\rangle$ is prepared with a quantum circuit in (a).}
\label{Sapply}
\end{figure}

The MBQC on the transformed resource state proceeds as follows.
(1) Initialize $|\psi_{\rm in}(n)\rangle|\psi_{\rm in}(1)\rangle|+i\rangle$ to $|0^{n+1}\rangle|+i\rangle$. This can be accomplished without the added $|+i\rangle$ because the given resource state is computationally universal, and $|\psi_{\rm in}(n)\rangle|\psi_{\rm in}(1)\rangle$ is a tensor product of $|0\rangle$'s and/or $|+\rangle$'s.
(2) Run the quantum circuit in Fig.~\ref{Sapply}(a) on the $(n+1)$th, $n$th, and $(n-1)$th input qubits $|000\rangle$. Therefore, the $(n+1)$th qubit becomes $|1\rangle$, which will be used to implement the $S$ gate in step 3 (for the proof, see Appendix B).
This step can also be implemented without the added $|+i\rangle$ because the given resource state is computationally universal.
(3) Depending on which quantum gate we want to apply on the first $n$ qubits, carry out one of the following procedures:
(a) When $H$ or $CCZ$ is applied, conduct the corresponding measurements in the original computationally universal MBQC.
(b) When $S$ is applied, run the quantum circuit in Fig.~\ref{Sapply}(b) by using $(n+1)$th and $(n+2)$th input qubits $|1\rangle|+i\rangle$ (for the proof, see Appendix B).
Note that $|1\rangle|+i\rangle$ is invariant in step 3; hence, the $S$ gate can be applied at any time.
In other words, $|1\rangle|+i\rangle$ can be used recursively.
(4) Finally, a desired output state is generated on the first $n$ qubits in ${\mathcal C}_O$ (up to a byproduct).
An advantage of our method is that the set of the required measurement bases does not change before and after our transformation.
This is because we use only $H$ and $CCZ$ gates in the above procedures.

{\it Application to MBQC with hypergraph states.}---Since the pattern of measurements depends on a resource state to which our catalytic transformation is applied, we cannot discuss its detail without specifying the resource state(, and hence we give our results by using quantum circuits in the previous section).
In this section, to state our results in terms of MBQC rather than the quantum circuit model, we apply our transformation to a concrete hypergraph state.
Hypergraph states are generalizations of graph states~\cite{RHBM13}.
Let $G\equiv(V,E_2,E_3)$ be a triplet of the set $V$ of $m$ vertices, the set $E_2$ of edges connecting two vertices, and the set $E_3$ of hyperedges connecting three vertices.
Note that hyperedges connecting more than three vertices are also generally allowed, but they are unnecessary in this section.
An $m$-qubit hypergraph state $|G\rangle$ corresponding to the hypergraph $G$ is defined as $(\prod_{(j,k,l)\in E_3}CCZ_{jkl})(\prod_{(j,k)\in E_2}\Lambda(Z)_{jk})|+\rangle^{\otimes m}$, where the subscript represents to which qubits the quantum gate is applied.

\begin{figure}[t]
\includegraphics[width=8cm, clip]{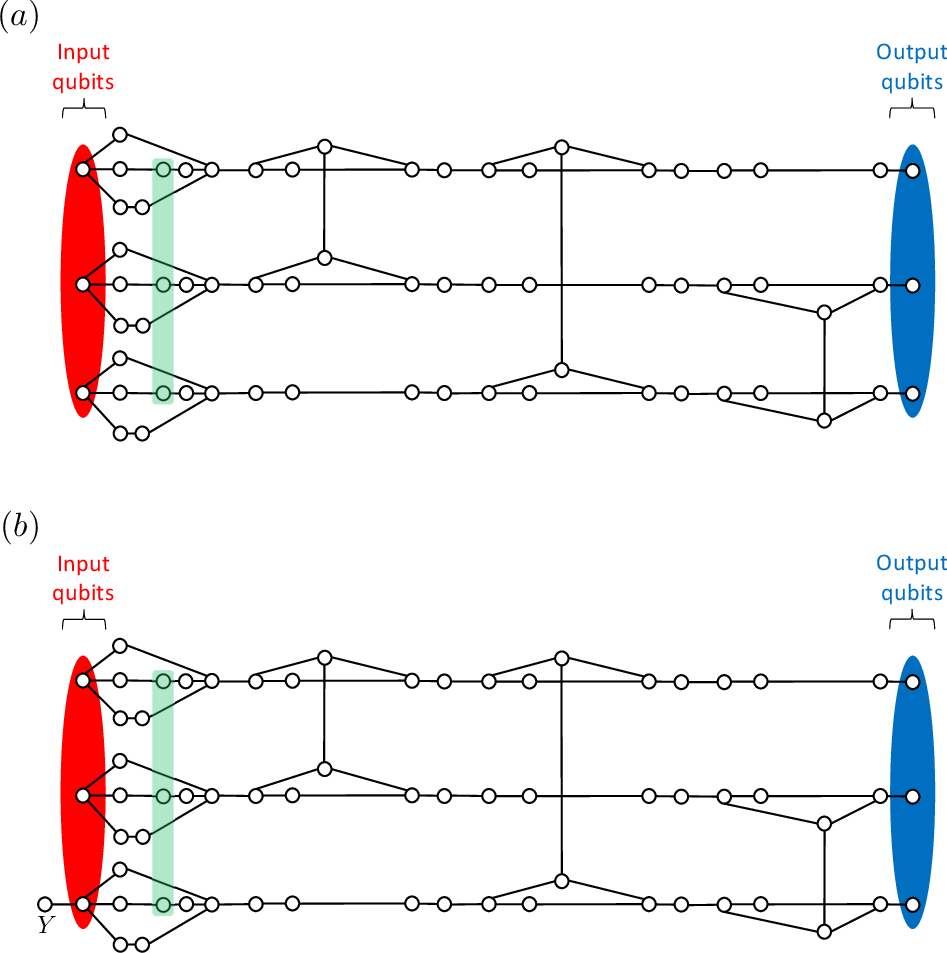}
\caption{Transformation of a computationally universal hypergraph state to a strictly universal one. Each vertex and edge represent $|+\rangle$ and the $CZ$ gate, respectively. Each green rectangle represents $CCZ$. (a) The hypergraph state $|G_3^1\rangle$ in Ref.~\cite{TMH19} with $n=3$ input qubits and depth $d=1$. Any $1$-depth quantum circuit composed of $\{H, CCZ\}$ can be realized on $|+\rangle^{\otimes 3}$ by measuring it in Pauli-$X$ and -$Z$ bases. By entangling a number of $|G_3^1\rangle$'s and $|+\rangle$'s using $CZ$ gates, the computationally universal hypergraph state~\cite{TMH19} is generated. (b) Transformed hypergraph state. Difference from (a) is that an additional single $|+\rangle$ is entangled with the third input qubit by using the $CZ$ gate. By measuring it in the Pauli-$Y$ basis, the third input qubit becomes $S|+\rangle=|+i\rangle$ (up to a byproduct) due to the gate teleportation.}
\label{hyper}
\end{figure}

The hypergraph state in Ref.~\cite{TMH19} was shown to be computationally universal and prepared by applying the controlled-$Z$ ($CZ$) gates to the $\Theta(n^4d)$ $|+\rangle$'s and $\Theta(n^3d)$ small hypergraph states shown in Fig.~\ref{hyper}(a), where $n$ and $d$ are the number of input qubits and depth under the gate set $\{H, CCZ\}$, respectively.
From our argument in the previous sections, the above hypergraph state~\cite{TMH19} can be made strictly universal by replacing an input qubit with a single $|+i\rangle$.
Such replacement can be accomplished by modifying one of the small hypergraph states such as that in Fig.~\ref{hyper}(b).
By measuring the added qubit in the Pauli-$Y$ basis, the third input state becomes $|+i\rangle$ (up to a byproduct), due to the gate teleportation.
Since the original hypergraph state~\cite{TMH19} requires only Pauli-$X$ and -$Z$ basis measurements for computational universality, the transformed hypergraph state achieves strict universality by using a single Pauli-$Y$ basis measurement in addition to those Pauli measurements (see also Appendix D).
Under the assumption that quantum computers are more powerful than classical computers, this Pauli universality cannot be obtained by using graph states because Pauli measurements on graph states can be efficiently simulated with a classical computer.
Note that the combination of the hypergraph states in Ref.~\cite{TMH19} and Pauli-$X$ and -$Z$ basis measurements cannot achieve the strict universality, i.e., cannot implement the $S$ gate.
This is because these states and measurements are real quantum states and operations, respectively.
Therefore, our transformation surely makes a nonstrictly universal MBQC to a strictly universal one.

Our strictly universal hypergraph state should be useful for distributed quantum computation~\cite{CCC20}, which is quantum computation conducted on quantum internet~\cite{FNN2}.
It is a promising approach to realize a universal quantum computer because it only requires small or intermediate-scale quantum computers to build a large-scale quantum computer.
To construct quantum internet, it would be important to convert from an entangled state to another one (e.g., from a graph state to the tensor product of Bell pairs~\cite{DHW20}).
Our hypergraph state can be used to prepare any entangled state (up to a byproduct) by just performing Pauli measurements in each quantum computer.
Furthermore, it is compatible with several quantum communication protocols such as quantum key distribution~\cite{BB14} and one-time programs~\cite{RKFW21} in the sense that they can also be implemented with Pauli measurements.
In Appendix D, to evaluate the practicality of this application, we discuss the required number of qubits.

{\it Conclusion and discussion.}---We have proposed a method of transforming from a computationally universal to strictly universal MBQC by simply replacing a single input qubit with a catalyst $|+i\rangle$~\cite{FNNN2}.
We believe our results will facilitate the discovery of novel strictly universal resource states.
By applying our transformation to the hypergraph state~\cite{TMH19}, we have constructed a strictly universal hypergraph state.
Our results in Fig.~\ref{hyper} should indicate that the gap between the computational and strict universalities is smaller than expected thus far.
In fact, our constructed strictly universal hypergraph state has the same amount of magic (i.e., nonstabilizerness) with the computationally universal hypergraph state~\cite{TMH19} when we quantify it by using the stabilizer rank~\cite{BSS16}.
It would be interesting to extensively explore the gap between computationally universal and strictly universal MBQCs with Pauli measurements from the viewpoint of magic (see, e.g., Ref.~\cite{LW22}).

In Ref.~\cite{KW19}, the strict universality of weighted graph states with Pauli-$X$ and -$Z$ basis measurements was shown.
Our hypergraph state also requires a Pauli-$Y$ basis measurement to achieve strict universality.
Although our results reduce the gap between them, weighted graph states are still slightly superior to hypergraph states.
However, with respect to verifiability (i.e., how easily the fidelity between the ideal state and an actual state can be estimated), hypergraph states are conversely superior to weighted graph states.
Although only Pauli-$X$ and -$Z$ basis measurements are sufficient for hypergraph states~\cite{TM18,ZH19}, non-Pauli measurements are required for weighted graph states~\cite{HT19}.
It would be interesting to investigate differences between them, which are two different generalizations of graph states, more deeply.
As another future work, it will be interesting to generalize our results to other nonstrictly universal gate sets.

\medskip
We thank Seiichiro Tani and Seiseki Akibue for their helpful discussions.
We also thank anonymous referees of Phys. Rev. Lett. for their insightful comments, which deepen the relation between our results and existing results and significantly clarify the usefulness of our results.
Y. T. is partially supported by the MEXT Quantum Leap Flagship Program (MEXT Q-LEAP) Grant No. JPMXS0120319794, JST [Moonshot R\&D -- MILLENNIA Program] Grant No. JPMJMS2061, and the Grant-in-Aid for Scientific Research (A) No. JP22H00522 of JSPS.

\clearpage
\widetext
\section{Appendix A: Proof of Fig.~\ref{CS}}
{\it Proof.} The equality in Fig.~\ref{CS} can be shown as follows.
The quantum circuit on the right of Fig.~\ref{CS} applies $HZ^{jk}HSHZ^{jk}H=X^{jk}SX^{jk}$ to the third qubit $|0\rangle$ when the state of the first and second qubits is $|jk\rangle$ with $j,k\in\{0,1\}$.
Here, $X\equiv|1\rangle\langle0|+|0\rangle\langle 1|$ is the Pauli-$X$ gate.
This means that $|0\rangle$ becomes $i|0\rangle$ if and only if $j=k=1$ and does not change in other cases.
It is trivial that for any $j$ and $k$, the state $|jk\rangle$ of the first and second qubits is invariant in this quantum circuit. 
Therefore, the quantum circuit on the right has the same function as the controlled-$S$ gate on the first and second qubits.
\hspace{\fill}$\blacksquare$

\section{Appendix B: Proof of Fig.~\ref{Sapply}}
{\it Proof.} We first show that the output state of the quantum circuit in Fig.~\ref{Sapply}(a) is $|100\rangle$.
From
\begin{eqnarray*}
[(I\otimes I\otimes H)CCZ]^4=(CCXCCZ)^2&=&[(I^{\otimes 2}-|11\rangle\langle 11|)\otimes I+|11\rangle\langle 11|\otimes (-iY)]^2\\
&=&(I^{\otimes 2}-|11\rangle\langle 11|)\otimes I+|11\rangle\langle 11|\otimes (-I)\\
&=&\Lambda(Z)\otimes I,
\end{eqnarray*}
where $CCX\equiv(I\otimes I\otimes H)CCZ(I\otimes I\otimes H)$ is the Toffoli gate and $Y\equiv i|1\rangle\langle 0|-i|0\rangle\langle 1|$ is the Pauli-$Y$ gate, the quantum circuit in the blue box can be considered as the controlled-$Z$ ($CZ$) gate.
Therefore, the output state of the quantum circuit in Fig.~\ref{Sapply}(a) is
\begin{eqnarray*}
\{(H\otimes H)[\Lambda(Z)(I\otimes H)]^2\Lambda(Z)(H\otimes H)\}|00\rangle\otimes|0\rangle&=&[(H\otimes I)\Lambda(X)\Lambda(Z)\Lambda(X)(H\otimes I)]|00\rangle\otimes|0\rangle\\
&=&[(H\otimes I)\Lambda(-Z)(H\otimes I)]|00\rangle\otimes |0\rangle\\
&=&[(H\otimes I)(Z\otimes I)(H\otimes I)]|00\rangle\otimes |0\rangle=|100\rangle.
\end{eqnarray*}
Note that $\Lambda(-Z)=Z\otimes I$ does not hold in general, but $\Lambda(-Z)|\psi\rangle|0\rangle=(Z\otimes I)|\psi\rangle|0\rangle$ holds for any single-qubit state $|\psi\rangle$.

Next, we show the Fig.~\ref{Sapply}(b).
Since the first input qubit is $|1\rangle$, all the $CCZ$ gates become the $CZ$ gates on the second and third qubits; hence, the first output qubit is trivially $|1\rangle$.
The unitary operation in the quantum circuit can be divided into the former and latter parts.
The latter part enclosed by the dotted red rectangle can be treated as
\begin{eqnarray*}
[(I\otimes H)\Lambda(Z)(I\otimes H)][(H\otimes I)\Lambda(Z)(H\otimes I)][(I\otimes H)\Lambda(Z)(I\otimes H)]=\Lambda(X)\tilde{\Lambda}(X)\Lambda(X)=SWAP
\end{eqnarray*}
on the second and third qubits, where $\tilde{\Lambda}(X)=I\otimes|0\rangle\langle 0|+X\otimes|1\rangle\langle 1|$ is the controlled-$X$ gate with swapped control and target qubits, and $SWAP\equiv\sum_{i,j\in\{0,1\}}|ji\rangle\langle ij|$ is the SWAP gate.
Therefore, the remaining task is to show that the output state of the former part is $S|\psi\rangle\otimes|+i\rangle$ for any single-qubit state $|\psi\rangle=\alpha|0\rangle+\beta|1\rangle$ with complex values $\alpha$ and $\beta$ satisfying $|\alpha|^2+|\beta|^2=1$.
This is shown by calculating as follows:
\begin{eqnarray*}
[\Lambda(Z)(H\otimes I)\Lambda(Z)(H\otimes H)\Lambda(Z)(I\otimes H)]|+i\rangle|\psi\rangle&=&(\Lambda(Z)\tilde{\Lambda}(X)\Lambda(X))|+i\rangle|\psi\rangle\\
&=&(\Lambda(Z)\tilde{\Lambda}(X))\cfrac{|0\rangle(\alpha|0\rangle+\beta|1\rangle)+i|1\rangle(\alpha|1\rangle+\beta|0\rangle)}{\sqrt{2}}\\
&=&\Lambda(Z)\cfrac{(\alpha|0\rangle+i\beta|1\rangle)|0\rangle+i(\alpha|0\rangle-i\beta|1\rangle)|1\rangle}{\sqrt{2}}\\
&=&S|\psi\rangle\otimes|+i\rangle.
\end{eqnarray*}
\hspace{\fill}$\blacksquare$

\section{Appendix C: Toward improving parallelizability}
\begin{figure}[t]
\includegraphics[width=7cm, clip]{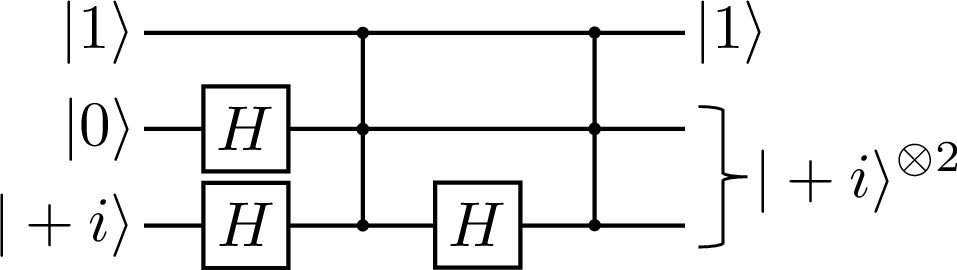}
\caption{Quantum circuit that duplicates $|+i\rangle$. Ancillary qubit $|1\rangle$ is prepared with quantum circuit in Fig.~\ref{Sapply}(a).}
\label{SS}
\end{figure}

As discussed in the main text, the $S$ gates cannot be applied in parallel in our method.
Toward relaxing this weakness, we propose a quantum circuit that duplicates $|+i\rangle$, as shown in Fig.~\ref{SS}.
Its correctness can be expressed as
\begin{eqnarray*}
CCZ(I\otimes I\otimes H)CCZ(I\otimes H\otimes H)(|10\rangle\otimes|+i\rangle)&=&\{I\otimes[\Lambda(Z)(I\otimes H)\Lambda(Z)(H\otimes H)]\}(|10\rangle\otimes|+i\rangle)\\
&=&[I\otimes(\Lambda(Z)\Lambda(X))](|1\rangle\otimes|+\rangle\otimes|+i\rangle)\\
&=&(I\otimes\Lambda(iY))(|1\rangle\otimes|+\rangle\otimes|+i\rangle)\\
&=&|1\rangle\otimes|+i\rangle^{\otimes 2},
\end{eqnarray*}
where $X\equiv|1\rangle\langle 0|+|0\rangle\langle 1|$ and $Y\equiv i|1\rangle\langle 0|-i|0\rangle\langle 1|$ are the Pauli-$X$ and -$Y$ gates, respectively.

\section{Appendix D: Remarks on application of catalytic transformation to hypergraph states}
In the main text, we construct a strictly universal hypergraph state by applying our catalytic transformation to the computationally universal hypergraph state in Ref.~\cite{TMH19}.
Since a single qubit is added due to the transformation (see Fig.~4(b) in the main text), the qubit number of the resultant strictly universal state is
\begin{eqnarray}
\label{RNQ1}
d(2n+63)\binom{n}{3}-n+1,
\end{eqnarray}
where $n$ and $d$ are the number of input qubits and depth under the gate set $\{H,CCZ\}$, respectively.
Even for the modest values of $n=6$ and $d=5$, the value of Eq.~(\ref{RNQ1}) becomes $7495$, which should be quit large for near-term quantum computers.
The reason of why the huge number of qubits are required for the hypergraph states in Ref.~\cite{TMH19} is that a single small hypergraph state in Fig.~4(a) is required to implement a single $CCZ$ gate, and we must consider $\binom{n}{3}$ kinds of $CCZ$ gates depending on which three input qubits are acted by the $CCZ$ gate.
These $\binom{n}{3}$ small hypergraph states are entangled in a straightforward manner by using $\Theta(n^4d)$ $|+\rangle$'s.
Fortunately, this issue can be solved by embedding a sorting network into MBQC as shown in Ref.~\cite{YFTTK20}.
More precisely, by applying our catalytic transformation to the computationally universal hypergraph state in Ref.~\cite{YFTTK20}, the required number of qubits is significantly decreased to at most
\begin{eqnarray}
\label{RNQ2}
3d\left[n\binom{\left\lceil\log_2{n}\right\rceil}{2}+2\left(2n-1\right)\right]+n+1,
\end{eqnarray}
which is just $607$ when $n=6$ and $d=5$.
Here, $\lceil\cdot\rceil$ is the ceiling function.
Note that the required measurement bases are still Pauli-$X$ and -$Z$ bases if we use the computationally universal hypergraph states in Ref.~\cite{YFTTK20}.
Thus, when we assume that we use the IBM Condor processor with $1121$ qubits~\cite{PJSO24}, the maximum achievable $n$ and $d$ are $26$ and $28$, respectively.
It is worth mentioning that the dependence of Eq.~(\ref{RNQ2}) on $n$ and $d$ seems to be optimal up to a poly-logarithmic factor.
This is because MBQC consumes at least a single qubit to implement a single quantum gate, and hence $\Omega(nd)$ qubits should be necessary to implement any depth-$d$ quantum computation on $n$ input qubits.

Someone may think that since the $S$ gate can be directly applied by measuring a qubit of any hypergraph state in the Pauli-$Y$ basis, it is redundant to use the quantum circuit in Fig.~3(b), which requires a certain depth.
However, the hypergraph state in Ref.~\cite{TMH19} is tailored for $\{H,CCZ\}$, and hence qubits measured to implement the $S$ gates are not included in it.
In other words, to directly implement the $S$ gates with Pauli-$Y$ basis measurements, we must drastically change the shape of the hypergraph state.
On the other hand, if we use our transformation (i.e., the quantum circuit in Fig.~3(b)), the shape is almost the same before and after the transformation as we can see from Fig.~4.
This simpleness is an advantage of our transformation.

\begin{thebibliography}{99}
\bibitem{FN1}There exist infinitely many computationally universal gate sets other than the commonly used gate set $\{H,CCZ\}$.
In this Letter, we focus on computationally universal gate sets that can precisely implement $H$ and $CCZ$, and our results can be applied to any such gate set, e.g., $\{\sqrt{H}, CCZ\}$ and $\{H, CCZ\}\cup R$ with $R$ being any set of quantum gates with elements that are real numbers in the Pauli-$Z$ basis.
\bibitem{FN3}Weighted graph states are generalizations of graph states. They are generated by applying $\prod_j\Lambda(R_z(\theta_j))$ to $|+\rangle$'s, where $R_z(\theta)\equiv|0\rangle\langle 0|+e^{i\theta}|1\rangle\langle 1|$ is the $Z$ rotation gate for any $\theta\in{\mathbb R}$. When each of $\{\theta_j\}_j$ is equal to $\pi$, weighted graph states become ordinary graph states.
\bibitem{FN2}In the MBQC on the correlation space~\cite{GE07}, the input state is defined as an edge state in the correlation space. Therefore, our transformation becomes the replacement of the edge state.
\bibitem{FNN2}As another potential application, we give the complexity of synthesizing quantum states (i.e., how hard to generate quantum states)~\cite{RY22,DGLM23,MY23,R24}.
This problem is well motivated by condensed matter physics whose purpose is to synthesize novel materials.
Considering that the complexity of synthesizing quantum states is a quantum analogue of computational complexity, which considers how hard to solve problems with classical inputs and outputs, and MBQC is a useful tool for several computational complexity classes such as ${\sf QMA}$~\cite{MHNF15,MNS16,MFN16} and ${\sf QIP}$~\cite{M16}, our strictly universal hypergraph state with Pauli measurements may give a fresh perspective to the hardness of generating quantum states.
\bibitem{FNNN2}The deformation from the star-lattice AKLT state, which is not expected to be universal, to a universal graph state was investigated in Ref.~\cite{DB14} to observe a phase transition in computational power.
Our study may also be used to observe the computational phase transition between two types of universality by gradually deforming an input qubit of a computationally universal resource state from the original state to $|+i\rangle$.
\end{thebibliography}
\end{document}